\documentstyle[epsfig]{mn} 

\begin{document}
 
\title
[ON THE RELATIONSHIP BETWEEN GALAXY FORMATION AND QUASAR EVOLUTION]
{ON THE RELATIONSHIP BETWEEN GALAXY FORMATION AND QUASAR EVOLUTION}

\author
[Alberto Franceschini, et al.]
{
Alberto Franceschini$^1$, Guenther Hasinger$^2$, Takamitsu Miyaji$^3$,
and Diego Malquori$^{1,4}$\\
\vspace*{1mm}\\
$^1$ Dipartimento di Astronomia, Universit\`a di Padova, 
Vicolo dell'Osservatorio 5, Padova, 35122, Italy. 
e-mail: franceschini@pd.astro.it
\\
$^2$ Astrophisikalisches Institut, An der Sternwarte 16, Postdam, 
14482, Germany 
\\
$^3$
Max-Planck-Inst. fuer Extraterrestrische Physik, 
Garching bei Muenchen, 85740, Germany
\\
$^4$
Institute d'Astrophysique de Paris, CNRS, 98bis Boulevard Arago,  
Paris, 75014, France
}

\date{Sep 14, 1999}
 
\maketitle
 
\begin{abstract} 

We compare the evolution with cosmic time of the star-formation
rate per comoving volume in galaxies and of the volume emissivity
by Active Galactic Nuclei, as a clue to understand the relationship
between black hole accretion and the formation of the surrounding
structure. We find an interesting similarity between the evolution 
rates for the total populations of galaxies and AGNs, 
which indicates that, on average, the history of BH accretion
tracks that of stellar formation in the hosting galaxies.
Similarly, the evolution of luminous quasars parallels that of the
stellar populations in massive spheroidal galaxies, in keeping with the
locally established association of supermassive BHs and galactic bulges.
We finally comment on our finding that high-luminosity, high-mass
systems evolve on a faster cosmic timescale than the lower mass ones:
to explain this, the theories of structure formation
based on the gravitational collapse of dark matter halos
have to be complemented with a detailed description of the dynamical
processes in the baryonic component, which dominate the formation 
and evolution in high-density environments.

\end{abstract} 
\vspace{5mm}

 
\section{INTRODUCTION}

For long time quasar evolution and galaxy formation have been seen
as quite unrelated items of observational cosmology. 
In addition to the difficulty to sample homogeneous redshift intervals
for the two source populations, the main reason for 
this dichotomy was a lack of understanding of the
physical processes ruling the quasar phenomenon and galaxy formation and
evolution.

In this context, an important discovery by the refurbished HST was that
most (if not all) massive galaxies in the local universe harbor a 
nuclear super-massive ($\simeq 10^7-10^9\ M_\odot$) dark object 
(probably a BH), with a mass proportional to that
of the hosting spheroid ($M_{BH}\simeq 0.002-0.005\ M_{spheroid}$;
Kormendy and Richstone 1995; Faber et al. 1997; Magorrian et al. 1998).
This is indeed the prediction of the canonical model for AGNs --
assuming energy production by gas accretion onto a super-massive
BH -- and is consistent with evolutionary schemes 
which interprete the quasar phase as a luminous 
short-lived event occurring in a large fraction of all normal
galaxies, rather than one concerning a small minority
of "pathological" systems (e.g. Cavaliere \& Padovani 1988).
This association of nuclear massive dark objects with luminous spheroids is 
also directly proven by HST out to moderate redshifts ($z\simeq 0.4$),
where the host galaxies of radio-quiet and radio-loud
quasars are found to be large massive ellipticals (e.g. Mc Lure et al. 1999).

On the distant-galaxy side, important progresses have been made
in the characterization of the
star-formation history as a function of redshift (Lilly et al. 1996;
Madau et al. 1996). Deep HST imaging also allowed to differentiate 
this history as a function of morphological type, for both cluster 
(Stanford et al. 1998) and field galaxies (Franceschini et al. 1998).

Interpretations of the relationship of quasar activity and 
galaxy formation have been attempted 
in the framework of the hierarchical dark-matter cosmogony
(e.g. Haehnelt, Natarajan \& Rees, 1998).
However, the QSO-galaxy connection is still enigmatic in several respects.
In particular, while hierarchical clustering seems to successfully account for
the onset of the quasar era at $z\simeq 3$ [assumed that supermassive BH form
proportionally to the forming dark matter halos], the progressive decay of the
quasar population at lower $z$ (the QSO evolution) is not accounted
for as naturally, and has to require a substantially more detailed
physical description (e.g. including the role of galaxy interactions, see 
Cavaliere \& Vittorini 1998).

To provide further constraints on this interpretative effort, we confront 
here the most detailed information on the evolutionary histories of the
AGN emissivity and of the stellar populations in galaxies.
The available information on the evolution of the Star Formation Rate (SFR) is
summarized in Sect. 2 , and compared in Sect. 3 with that of AGN emissivity.
Some consequencies, in particular concerning the stellar and
BH remnants of the past activity and implications for models of structure
formation, are discussed in Sect. 4, while Sect. 5 summarizes our
conclusions.
$H_0=50\ Km/s/Mpc$ and two values ($q_0$=0.5 and 0.15) for the deceleration
parameter are assumed.

\section{STELLAR FORMATION HISTORIES OF GALAXY POPULATIONS}

The systematic use of the Lyman "drop-out" technique 
to detect $z>2$ galaxies (Steidel et al. 1994),
combined with spectroscopic surveys of galaxies at $z\leq 1$ (Lilly et al.1995)
and $H\alpha$-based measurements of the present-day SFR, 
have allowed to determine the evolution of the SFR in galaxies over most of the
Hubble time (Madau et al. 1996).
Figures 1 and 2 summarize some results of these analyses (for two values of 
$q_0$) in the form
of small filled circles, where the SFR is measured from fits
of the UV rest-frame flux (assuming
a Salpeter IMF with low mass cutoff $M_{low}=0.1\ M_\odot$).  
We see, in particular, an increase by a factor 
$\sim 10-20$ of the SFR density going from z=0 to $z\simeq 1-1.5$.

The main uncertainty in these estimates is due to the - a priory unknown -
effect of dust, erasing the optical-UV flux and 
plaguing the identification of distant reddened galaxies in the "drop-out"
catalogues. Corrections for this effect have been estimated to range from 1 to 3
magnitudes, and mostly apply above $z\sim 1$. Although rather uncertain at the
moment, the SFR density above redshift 1 seems to keep roughly constant up to
z$\simeq 3$ (the datapoint at z=3 in Fig. 1 comes from Meurer et al. 1999).

Exploiting the exceptional imaging quality and spectral coverage
available in the Hubble Deep Field, 
Franceschini et al. (1998) have analysed a sample of
morphologically-selected early-type galaxies with $z<1.3$.
By fitting synthetic galaxy spectra to these data, it has been possible to 
estimate
the baryonic mass and age distribution of stellar populations, and hence
the star-formation history per comoving volume (reported as shadowed 
regions in Figs. 1 and 2):
the evolutionary SFR density of field early-type galaxies is roughly constant
at $z\geq 1.5$, while showing a fast convergence at lower redshifts
(correspondingly, 90\% of stars in E/S0 galaxies 
have been formed between $z\simeq 1$ and 3 for $q_0=0.15$;
$z\simeq 1$ and 4 for $q_0=0.5$). This evolution pattern is quite different from
that of the general field population (small circles in Figs. 1 and 2),
which displays a much shallower dependence on cosmic time.

We note that at $z>2$ the uncertainties in all estimates
of the SFR become very serious. The SFR estimate of early-type galaxies 
by Franceschini et al. is based on fits to the SEDs of galaxies
at $z<1.3$, and then become uncertain at $z>2$. Similar problems affect
the SFR estimates for the Lyman "drop-out" galaxies, because of the reddening
problem.
The present analysis will therefore mostly concentrate on data at $z<2$.

Thus {\it galaxies with early morphological
types -- dominating the galaxy populations in high-density environments --
show an accelerated formation history, while later types, more typical
of low-density regions, are slower in forming their stellar content}.
In the merging picture of elliptical and S0 galaxy formation
(e.g. Toomre \& Toomre 1972; Barnes \& Hernquist 1992)
an accelerated SF history in high-density environments
is naturally accounted for by a higher rate of
galaxy interactions and mergers at early epochs.

\section{FORMATION AND EVOLUTION OF ACTIVE GALACTIC NUCLEI}

Quasars and AGNs have been studied in all accessible wavebands,
in all of them showing evidence for strong cosmological evolution.
The most efficient AGN selection comes from X-ray surveys:
a purely flux-limited sample at X-ray energies above
0.5 KeV contains a vast majority of AGNs with no particular
bias as a function of redshift. The same is not true for optical searches, 
suffering incompleteness at both low-luminosities
(because of the contribution of the host galaxy reddening the colours)
and at the high redshifts (because of the effect of dust and intergalactic
opacity), and for surveys of radio sources, since only a minority of AGNs 
emit in the radio.
A further crucial advantage of the X-ray selection is provided
by the X-ray background (XRB), which sets an integral constraint on AGN 
emissivity at fluxes much fainter than observable with the present 
facilities.

\epsfxsize=9cm
\begin{figure}[!Ht]
\vspace*{-10pt}
\hspace*{0pt}
\epsffile{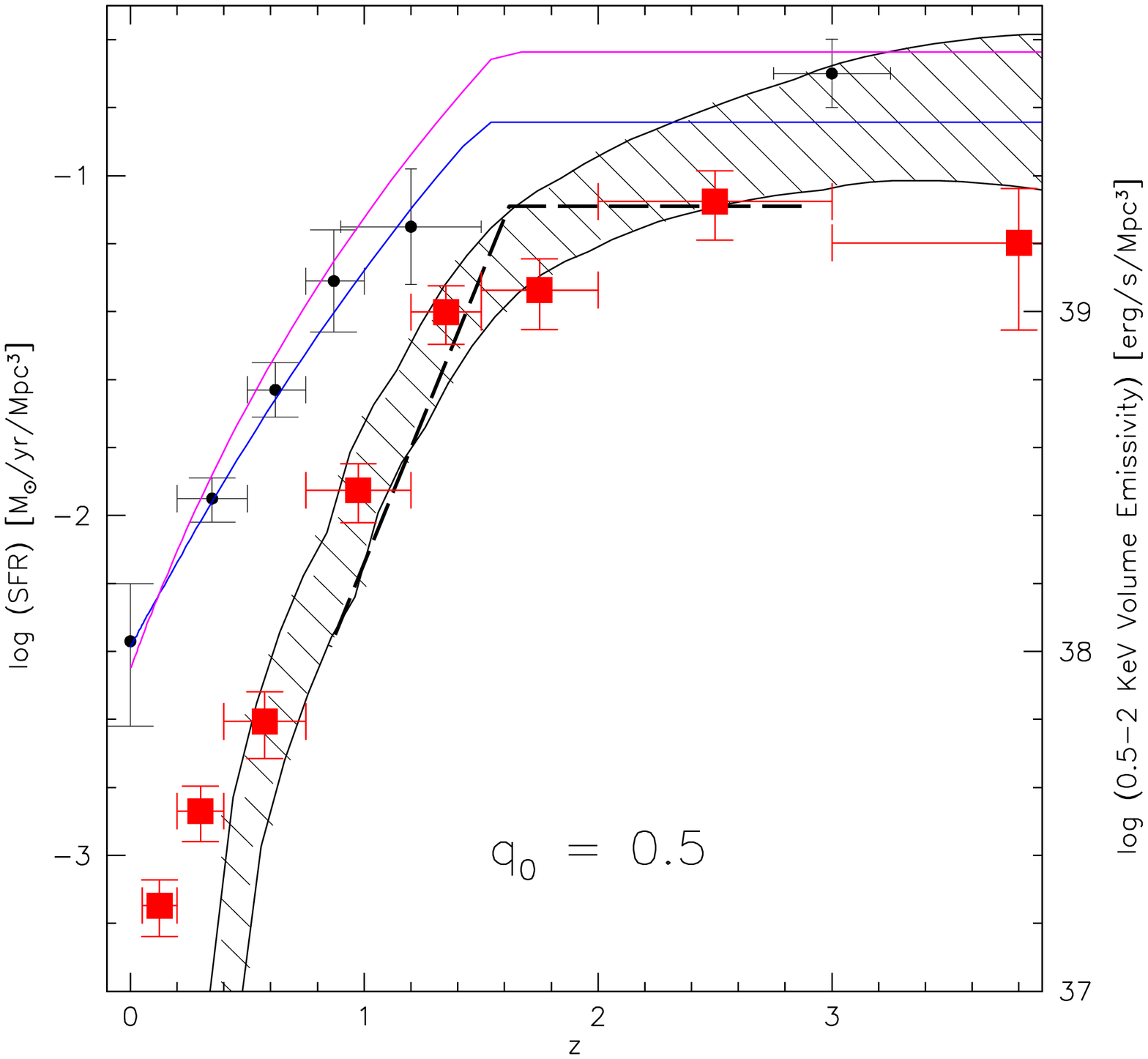}
\vspace*{-10pt}
\caption {Comparison of the redshift evolution of AGN emissivity 
per comoving volume (to be read on the right-hand axis) 
with the evolution of the star-formation rate 
in galaxies (left-hand axis), for $q_0=0.5$. 
Small filled circles are the evolutionary SFR of field galaxies
estimated from
conversion of the rest-frame optical-UV flux to SFR.
The shaded region describes the evolution of the SFR in field
galaxies morphologically classified as ellipticals and S0s by
Franceschini et al. (1998).
The filled squares are the 0.5 to 2 keV comoving volume emissivities
(in $erg/s/Mpc^3$) of high-luminosity AGNs 
($L_{0.5-2 keV}\geq 10^{44.25}\ erg/s$, see Table 1).
The top two lines are the 0.5-2 keV volume emissivities of the 
total AGN population (normalized by the same factor), based on the
LDDE1 and LDDE2 models producing 60\% and 90\% of the XBR at 1 KeV,
respectively. The thick dashed line is the volume emissivity
of optical quasars from Schmidt, Schneider \& Gunn (1995).
The scales of the two vertical axes are chosen in such a way that
AGN volume emissivity and SFR density overplot each other.
}
\label{fig1}
\end{figure}

Miyaji et al. (1999) have used a large X-ray AGN database
of 7 ROSAT samples -- including wide-area surveys (Appenzeller et al. 1998) 
and very deep PSPC and HRI integrations
on small selected areas (Hasinger et al. 1998) -- to analyse the evolution
of X-ray AGNs over large redshift ($0<z<5$) and luminosity
($10^{42}<L_{0.5-2 keV}<10^{47}\ erg/s$) intervals.
Various kinds of evolution patterns have been tested, 
including Pure Density (PDE), Pure Luminosity 
(PLE), and Luminosity Dependent Density Evolution (LDDE).
These X-ray surveys indicate for the first time 
that the AGN evolution is inconsistent with
strict number density conservation (PLE), and rather
require a combined increase with redshift of the source number
and luminosity (LDDE).

The X-ray luminosity function $n(L_{x},z)$ (sources per
unit volume and unit logarithmic $L_x$ interval) 
is modelled by Miyaji et al. as a two power-law expression
$$
 n(L_{x},z) \propto [(L_{x}/L_\ast)^{\gamma_1}+
(L_{x}/L_\ast)^{\gamma_2}]^{-1} \cdot e(L_{x},z),
$$
where $L_x$ the luminosity in the 0.5-2 keV band,
$e(L_{x},z)$ is the luminosity-dependent
evolution term and $\gamma_1\simeq 0.6,\ \gamma_2\simeq 2.3$. 
The best-fit LDDE model implies a lower
evolution rate at lower $L_x$, and simple density evolution (PDE) 
for the higher luminosity sources:
$e(L_{x},z)=(1+z)^{p}$, where $p$ linearly increases with $log L_x$
for $L_x\leq 10^{44-44.5}\ erg/s$ and is $\simeq 5.5$ above.
The number density of high-$L_x$ QSOs as a function of $z$
has been used by Hasinger (1998) to estimate the evolution of the AGN 
emissivity.

We report in Table 1 and in Figs. 1 and 2 (as large filled squares)
estimates of the soft X-ray emissivity per comoving volume 
versus $z$ for AGNs more luminous than $L_{th}= 10^{44.25}\ erg/s$
(this is the luminosity threshold above which a pure density
evolution applies).
It is apparent in these plots a {\it remarkable similarity
in the $z$-dependence of the emissivity of high-luminosity
AGNs and the SFR history of elliptical/S0 galaxies (shadowed
areas). Both functions show a quite faster decrease with 
cosmic time of the emissivity with respect to
that of the general field population.}
A very steeply declining emissivity of luminous AGNs is also confirmed by the
fast evolution of optical quasars (thick dashed lines).

This result is in agreement with the local evidence for an association between
nuclear super-massive dark objects and massive spheroids,  and with the
identification of the hosts of luminous quasars as galaxies with an early-type
morphology.

\epsfxsize=9cm
\begin{figure}[!Ht]
\vspace*{-10pt}
\hspace*{0pt}
\epsffile{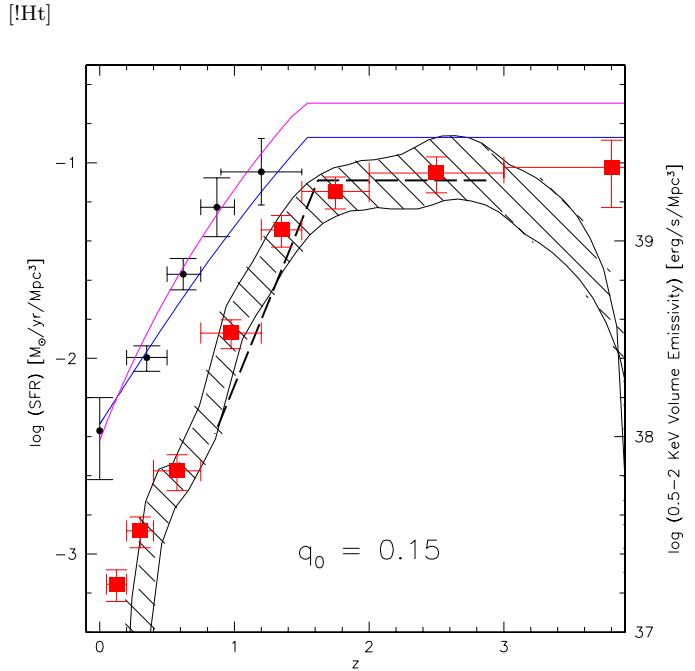}
\vspace*{-10pt}
\caption {Same as in Figure 1, for $q_0=0.15$
(SFR densities to be red on the
left y-axis, AGN volume emissivities on the right y-axis).
Small filled circles at $z\leq 1.2$ are the galaxy SFR transformed 
from $q_0=0.5$ to $q_0=0.15$ by correcting luminosities and 
volumes within each redshift interval (correction factors
being 1, 0.9, 1.15, 1.21 and 1.27 for data bins ranging
from z=0 to z=1.2).
}
\label{fig2}
\end{figure}

On the other end, a direct comparison with the population 
emissivities of lower luminosity AGNs is difficult, as
the present sensitivity limits prevent measurement of the LF of very 
faint AGNs. Here the information from the XRB becomes essential.
Adopting the LDDE best-fit model as a
description the LF of medium- to high-L AGNs, we  
extrapolated it to lower $L_x$ in such a 
way to reproduce most of the observed low-energy XRB.
The top two continuous lines in Figs. 1 and 2 are
the AGN 0.5-2 KeV luminosity density as
a function of $z$ corresponding to such two extrapolations,
one producing 60\% (henceforth LDDE1) and the other 
90\% (LDDE2) of the XRB at 1 KeV.

We see an accurate match between the redshift dependences
of the global AGN emissivity and of the rate of formation
of stars in the field galaxy population at $z<1.2$.
At $z\geq 1.5$ X-ray observations indicate 
that the AGN emissivity keeps constant, roughly as it has been 
found for the SFR density. Then,
{\it within the uncertainties, the close association of galaxy and AGN 
evolution may hold over a substantial redshift interval.}

{\small
\begin{verbatim}
_________________________________________________________
TABLE 1.  Luminosity density Ld between 0.5 and 2 keV per
	  comoving volume in erg/s/Mpc^3 of AGN with
	  log L>log L_th=44.25. Note that for any values 
	  of log L_th>44, Ld scales proportionally to
	  (L_th/10^44.25)^1.3
		  
 zmin  zmax       Ld    sig(Ld)         Ld    sig(Ld)
               ----q0=0.5------      ----q0=0.15-----      
 0.05  0.20    1.71e37  3.20e36      1.68e37  3.08e36
 0.20  0.40    3.24e37  6.02e36      3.15e37  5.66e36 
 0.40  0.75    5.95e37  1.33e37      6.39e37  1.33e37 
 0.75  1.20    2.85e38  5.58e37      3.23e38  5.39e37 
 1.20  1.50    9.52e38  1.87e38      1.09e39  2.02e38
 1.50  2.00    1.10e39  2.60e38      1.71e39  3.18e38  
 2.00  3.00    2.01e39  4.62e38      2.13e39  4.43e38 
 3.00  4.60    1.52e39  6.79e38      2.27e39  8.56e38
________________________________________________________
\end{verbatim}
  }  

How significant is the difference in the evolution rates between
high and low X-ray luminosity AGNs? Assuming a luminosity threshold at
$L_{th}=10^{44.25}\ erg/s$, we find:

a) the LDDE1 $q_0=0.15$ model implies an increase in the volume 
emissivity going from $z=0$ to $z=1.5$ by a factor 100 for sources
with $L>L_{th}$ and by a factor 10 for sources with $L<L_{th}$; 
in the $q_0=0.5$ case, the two factors become 100 and 16;

b) for the LDDE2 $q_0=0.15$ model the
evolution in the volume emissivity is by a factor 100 for $L>L_{th}$
and 30 for $L<L_{th}$, while for $q_0=0.5$ the two factors become
100 and 50.

Altogether, there is at least a factor 2 difference in the evolution
rates of high-$L_x$ and low-$L_x$ AGNs, which may get as high as 10 for the
open LDDE1 case.

\section{DISCUSSION}

The similarities of the evolution rates between high-L AGNs and massive
spheroidal galaxies, as well as those between low-L AGNs and the field galaxy
population, call for a close relationship of the processes triggering
star-formation with those responsible for the gas flow fueling the massive BH.
Were the SF and BH accretion processes really concomitant, we would expect that
the luminosity functions of starbursting galaxies and of quasars and AGNs should
have similar shapes and perhaps similar normalizations when the luminosities are
expressed in bolometric units. This seems indeed indicated by
the comparative study of the bolometric luminosity functions of luminous
far-IR galaxies (the far-IR flux being a good measure of the SFR),
of bright optical quasars (Schmidt \& Green 1983), and lower luminosity
Markarian Seyferts, as reported by Soifer et al.
(1987) and Sanders \& Mirabel (1996): these LFs have been found to share
not only very similar power-law shapes at the high luminosities 
($n[L_{bol}]\sim L^{-2.1}$), but even similar 
(within a factor of $\sim 2$) normalizations.

We interprete this coincidence as a further support to our results.

\subsection{The energetics associated with SF and AGN activity}

Although substantial uncertainties are inherent in
the absolute normalization of various curves in Figs. 1 and 2, 
it may be worth making an order-of-magnitude comparison
of the energetics associated with the processes of stellar formation 
in galaxies and BH accretion in AGNs. Interesting constraints will
ensue from matching them with the local remnants (i.e. low-mass stars and
supermassive BHs) of the past activities.

Let us first evaluate the global bolometric emissivity by AGNs as a function of
$z$ by applying a bolometric correction to the 0.5-2 KeV emissivity (top two
lines in the figures).
From the average X-ray to optical spectral index 
$\alpha_{ox}=log(L_{2500 A}/L_{2 KeV})/2.605\simeq 1.4$
(La Franca et al. 1995) we obtain the $2500 A$ flux, and then correct
it by a further factor 5.6 to get the bolometric flux (Elvis et al. 1994).
For type-I AGNs (dominating the soft X-ray samples) we then find
%
$L_{BOL, type I}\simeq 50\ L_{0.5-2 keV}.$
%
Since type-I AGNs contribute only $\sim 20\%$ to the
total hard X-ray background (e.g. Schmidt \& Green 1983), we have another
factor $\sim 5$ to account for the contribution of type-II AGNs to the global
emissivity:
\begin{equation}
L_{BOL, AGN}\sim 5\ L_{BOL, type I}\sim 250 \left({f_{BOL} \over 250}\right)\ 
L_{0.5-2 keV},
\end{equation}
where $f_{BOL}$ parametrizes the average bolometric correction factor
to the 0.5-2 keV flux.

As for galaxies, the scaling factor from $\dot M\ (M_\odot/yr)$
to bolometric luminosity is simply given by
\begin{equation}
 L_{BOL, SFR}\ [erg/s] = c^2\epsilon \dot M \sim 6\ 10^{43}
{\epsilon \over 0.001}\ {\dot M\ \over [M_\odot/yr]},$$
\end{equation}
for a stellar radiative efficiency of $\epsilon=0.001$ 
consistent with the assumption by Franceschini et al. (1998) 
and Madau et al. (1996) of a Salpeter IMF 
with lower mass limit of $M_{low}=0.1\ M_\odot$ and primordial
initial composition (newly born stars are assumed to immediately 
release most of their energy).

Assuming a fiducial $\eta=10\%$ radiation efficiency by BH accretion,
and taking into account that there is a factor $2.4\ 10^{40}$ to bring
the left-hand axis scale in Figs. 1 \& 2 to coincide with that of the
right-hand axis, we find that the ratio of the mass $M_{BH}$ of the
remnant locked as a supermassive BH after the 
AGN phase to the remnant in low-mass stars ($M_\ast$) should be:
\begin{equation}
 M_{BH} \simeq 0.001 \left({\epsilon \over 0.001}\right) 
\left({0.1 \over \eta}\right) \left({f_{BOL} \over 250}\right) \ M_\ast .
\end{equation}
If compared with the locally observed ratio of $M_{BH}$ to the mass 
of the hosting spheroid
\begin{equation}
 M_{BH} \simeq (0.002-0.005)\ M_{spheroid},
\end{equation}
this result is consistent with the fact that nuclear massive BH's
are found primarily associated with galactic bulges, and not e.g. with
the disk components which substantially contribute to $M_\ast$ in eq.(3).

A similar exercise comparing the low-mass stellar remnants in E/S0's
with the BH remnants of massive/luminous quasars (respectively
shaded regions and filled squares in the figures) would involve definition 
of the X-ray luminosity threshold $L_{th}$ above which AGN are hosted by 
massive spheroidal galaxies, and the ratio of type-II to type-I quasars.
Assuming as above $L_{th}=10^{44.25}$ would bring to a relation
between the BH mass and the mass of the hosting spheroid 
similar to that in eq.(3). If we consider that the ratio of type-II to
type-I objects among luminous quasars is lower than for lower luminosity 
AGNs, then to get consistency with the large observed ratio of eq.[4] would 
require either:
a) a very low radiative efficiency in AGNs ($\eta<0.1$, e.g. due to a violent
super-Eddington accretion phase), or b) an
higher bolometric correction for AGNs $f_{BOL}>250$ (e.g. because of a heavily 
dust-extinguished phase during quasar formation; Haehnaelt et al. 1998, Fabian
\& Iwasawa 1999), or finally, c) a very high ($\epsilon > 0.001$) radiative
efficiency of stars in spheroidal galaxies, as allowed by top-heavy stellar
IMFs (e.g., for $M_{low}=8M_\odot$, $\epsilon$ in eq.[3]
would increase to $\sim 0.006$).
Note that lowering the value for $L_{th}$, while improving the match of the
predicted BH mass in eq.(3) with the observation in eq.(4), would also 
spoil unacceptably the match between the evolution rates of QSO's and galaxies
in Figs. 1 and 2, hence is not a solution.

\subsection{Quasars and models of structure formation}

We discussed evidence that luminous quasars, associated with massive
BH's in massive spheroidal galaxies, evolve 
on a shorter cosmic timescale than
lower luminosity, lower mass objects. This result does not seem to 
fit into simple predictions based on the
gravitationally-driven Press-Schechter formalism.
For example Haiman \& Menou (1998) obtain from
the Press-Schechter theory a much faster decay 
with cosmic time of the AGN
accretion rate (hence of the AGN luminosity) for
low-mass BH's in low-mass dark-matter halos,
while that of massive objects is barely expected to decrease
from z=3 to z=0. In fact, quite the opposite trend is indicated by our
analysis in the previous Sections.
A similar problem likely arises when explaining with a process of purely 
gravitational clustering the origin of galaxies, and of their spheroidal 
components in particular.

What is the trigger and the ruling mechanism for 
generating a galaxy spheroid together
with a super-massive collapsed remnant including $\sim$0.2\%
of the mass of the host? 
As discussed by many authors (Kormendy \& Sanders 1992;
Barnes \& Hernquist 1992), {\it a ruling effect is in the gas/stellar 
dynamical processes related with galaxy interactions 
and mergers. This is probably the only way to achieve
stellar systems with very high central concentrations
as observed in early-type galaxies,
an obviously favourable birth-place for massive
nuclear star-clusters and a super-massive collapsed object.}

The merging/interaction concept, together with
the progressive exhaustion of the fuel, 
allows to understand the "accelerated" evolution of 
luminous AGNs (Cavaliere \& Vittorini 1998), but also the fast
decay with time of the SF in massive spheroidal galaxies. 
In cosmic environments with higher-than-average
density the forming galaxies have experienced a
high rate of interactions already at high
redshifts ($z>1$), this bringing shortly to a population
of spheroid-dominated galaxies (observable at low-z
mostly in groups and clusters of galaxies) 
containing a massive BH whose accretion history is the
same as that of the host galaxy.
In lower density environments the interaction rate
was slower, and proportionally lower was the mass
locked in a bulge with 
respect to that making stars quiescently in a disk.
Galaxies in these low-density environments keep forming stars 
and accreting matter on the BH down to the present epoch and 
form the low-redshift AGN population.

\section{CONCLUSIONS}

We have found a close match bewteen the evolution rates of 
the star-formation in galaxies and of the volume emissivity
in AGNs. This similarity seems to hold not only for
high luminosity AGNs ($L_{x}> 10^{44.25}\ erg/s$) compared with
the SF in massive spheroidal galaxies, but also when
comparing the average properties of the global populations, including
low mass/luminosity systems. 
A concomitance of SF and AGN/quasar activity is also locally indicated
by the almost coincidence of the bolometric luminosity functions
of luminous star-forming IR galaxies with those of optical quasars 
and Markarian Seyferts.
Then the same processes triggering the formation of stars also 
make a fraction of the available gas to accrete and fuel the AGN.
These processes are likely to be the interaction and merging events
between gas-rich
systems (Barnes and Hernquist 1992; Cavaliere and Vittorini 1998).

Assuming standard energy production efficiencies for BH accretion and
star formation, we have found rough agreement between the volume
emissivities by distant quasars and star-forming galaxies and the
observed ratios of low-mass stellar and supermassive BH remnants 
in local objects.

We have finally compared our finding that high mass/luminosity systems evolve
on a faster cosmic timescale than the lower-mass ones with predictions
based on structure formation theories based on the gravitational clustering
and coalescence of dark matter halos. Altogether, if the latter provide
the needed background conditions for the development of structures, 
much more physics is needed
-- in terms of gas/stellar dynamical processes  related with 
galaxy interactions, in terms of the progressive exhaustion of the
baryonic fuel available and of feedback reaction -- 
to explain observations of AGN evolution and (spheroidal) galaxy 
formation.

\centerline{\bf Acknowledgements}

We are indebted to an anonymous referee for useful comments and criticisms,
and to A. Cavaliere for discussions. Work partly supported by 
ASI Grant ARS-96-74 and the EC TMR Networks 'Galaxy Formation' and
'ISO Surveys'.

{}


\end{document}